\documentclass[twocolumn,showpacs,unsortedaddress,superscriptaddress,pra,aps,10pt,floatfix]{revtex4-1}
\usepackage {graphicx}
\usepackage {tabularx}
\usepackage{epsf}
\usepackage{braket}
\usepackage{array}

\begin{document}

\title{Electron shakeoff following the $\beta^+$ decay of  $^{19}$Ne$^+$ and  $^{35}$Ar$^+$  trapped ions}
%%%%%%%%%%%%%%%%%%%%%%%%%%%%%%%%%%%%%%%%%%%%%%%%%%%%
%%%%%%%%%%%%%%%%%%%%%%%%%%%%%%%%%%%%%%%%%%%%%%%%%%%%%
\author{X.~Fabian}
\altaffiliation[Present address:]{ Univ Lyon, Université Lyon 1, CNRS/IN2P3, IPN-Lyon, F-69622, Villeurbanne, France}
\affiliation{Normandie Univ, ENSICAEN, UNICAEN, CNRS/IN2P3, LPC Caen, 14000 Caen, France}
\author{X.~Fl\'echard}
\email{flechard@lpccaen.in2p3.fr}
\affiliation{Normandie Univ, ENSICAEN, UNICAEN, CNRS/IN2P3, LPC Caen, 14000 Caen, France}
\author{B.~Pons}
\affiliation{CELIA, Univ. Bordeaux - CNRS UMR 5107 - CEA, F-33405 Talence,  France}
\author{E.~Li\'enard}
\author{G.~Ban}
\affiliation{Normandie Univ, ENSICAEN, UNICAEN, CNRS/IN2P3, LPC Caen, 14000 Caen, France}
\author{M.~Breitenfeldt}
\altaffiliation[Present address:]{ PH Department, CERN, CH-1211 Geneva 23, Switzerland}
\affiliation{Instituut voor Kern- en Stralingsfysica, Katholieke Universiteit Leuven, B-3001 Leuven, Belgium}
\author{C.~Couratin}
\affiliation{Normandie Univ, ENSICAEN, UNICAEN, CNRS/IN2P3, LPC Caen, 14000 Caen, France}
\author{P.~Delahaye}
\affiliation{GANIL, CEA/DSM-CNRS/IN2P3, Caen, France}
\author{D.~Durand}
\affiliation{Normandie Univ, ENSICAEN, UNICAEN, CNRS/IN2P3, LPC Caen, 14000 Caen, France}
\author{P.~Finlay}
\affiliation{Instituut voor Kern- en Stralingsfysica, Katholieke Universiteit Leuven, B-3001 Leuven, Belgium}
\author{B. Guillon}
\author{Y. Lemi\`ere}
\author{F. Mauger}
\affiliation{Normandie Univ, ENSICAEN, UNICAEN, CNRS/IN2P3, LPC Caen, 14000 Caen, France}
\author{A.~M\'ery}
\affiliation{CIMAP, CEA-CNRS-ENSICAEN-UNICAEN, Normandie Univ, 14050 Caen, France}
\author{O.~Naviliat-Cuncic}
\affiliation{Normandie Univ, ENSICAEN, UNICAEN, CNRS/IN2P3, LPC Caen, 14000 Caen, France}
\affiliation{National Superconducting Cyclotron Laboratory and Department of Physics and Astronomy, Michigan State University, East-Lansing, MI, USA}
\author{T.~Porobic}
\affiliation{Instituut voor Kern- en Stralingsfysica, Katholieke Universiteit Leuven, B-3001 Leuven, Belgium}
\author{G.~Qu\'em\'ener}
\affiliation{Normandie Univ, ENSICAEN, UNICAEN, CNRS/IN2P3, LPC Caen, 14000 Caen, France}
\author{N.~Severijns}
\affiliation{Instituut voor Kern- en Stralingsfysica, Katholieke Universiteit Leuven, B-3001 Leuven, Belgium}
\author{J-C.~Thomas}
\affiliation{GANIL, CEA/DSM-CNRS/IN2P3, Caen, France}
%%%%%%%%%%%%%%%%%%%%%%%%%%%%%%%%%%%%%%%%%%%%%%%%%%%%%
\date{\today}
%%%%%%%%%%%%%%%%%%%%%%%%%%%%%%%%%%%%%%%%%%%%%%%%%%%%%

\begin{abstract}
The electron shakeoff of $^{19}$F and $^{35}$Cl atoms resulting from the $\beta$$^+$ decay of  $^{19}$Ne$^+$ and $^{35}$Ar$^+$ ions has been investigated using a Paul trap coupled to a time of flight recoil-ion spectrometer. The charge-state distributions of the recoiling daughter nuclei were compared to theoretical calculations based on the sudden approximation and accounting for subsequent Auger processes. The excellent agreement obtained for $^{35}$Cl is not reproduced in $^{19}$F. The shortcoming is attributed to the inaccuracy of the Independent Particle Model employed to calculate the primary shakeoff probabilities in systems with rather low atomic numbers. This calls for more elaborate 
calculations, including explicitly the electron-electron correlations.
\end{abstract}

 %%%%%%%%%%%%%%%%%%%%%%%%%%%%%%%%%%%%%%%%%%%%%%%%%
\maketitle
\section{Introduction}
\label{Sec_Intro}
Precision measurements in nuclear $\beta$ decay constitute sensitive probes to test the standard model of elementary particles. They provide clean and efficient means to search for new physics such as the existence of exotic couplings, time reversal violation, or a non unitarity of the Cabibbo-Kobayashi-Maskawa coupling matrix \cite{ Holstein14,Severijns14}. In particular, measurements of the recoil-ion energy distribution in the decay of well selected $\beta$ emitters were used to establish the vector - axial-vector structure of the weak interaction \cite{Hamilton47,Allen59,Johnson63}. They give access to the so called $\beta-\nu$ angular correlation coefficient which is sensitive to scalar and tensor exotic couplings excluded by the Standard Model of elementary particles \cite{Severijns14}. 

During the last two decades, the search for such exotic interactions has motivated new experiments based on modern ion-trapping or atom-trapping techniques coupled to intense radioactive beams \cite{Gorelov05,Vetter08,Flechard11,Ban13,Sternberg15,Hong16}. The most recent or ongoing experiments detect the $\beta$ particles and the recoil-ions in coincidence, providing a precise recoil-ion energy measurement using time of flight (TOF) techniques. The use of an electric field, to achieve maximum collection efficiency of the recoil-ions, makes these measurements also sensitive to the final charge-state of the recoiling daughter. Beside probing the true nature of the weak interaction, such experiments can thus also be extremely useful to investigate atomic processes induced by nuclear processes. 

Electron shakeoff (SO) resulting from the sudden change of the central potential is one of these fundamental atomic processes that can be addressed through the simple measurement of the charge-state distribution of the recoiling ions. The final charge distribution is, in a first step, the consequence of a primary ionization process caused by the central potential change, the sudden recoil of the daughter nucleus, and direct collision with the beta particle. This primary process can then possibly induce further ionization in terms of subsequent Auger electron emission. It is commonly accepted that the contribution of a direct knock out of a bound electron by the emitted beta particle is very small~\cite{Freedman74}, due to the large mismatch between the electron binding energy, typically less than 1 keV, and a beta particle kinetic energy of a few MeV in most cases. The dominant primary ionization process, the electron SO, is thus caused by the rapid change of the nuclear charge and, to some extent, by the sudden recoil velocity acquired by the daughter nucleus. Since the $\beta$ decay process is very rapid, of the order of 10$^{-18}$ s, SO ionization probabilities can be conveniently calculated in the framework of the sudden approximation (SA).

 In the beta decay of $^{6}$He$^+$ ions, an ideal textbook case with only one electron, simple quantum calculations based on the sudden approximation were tested with a relative precision smaller than $4{\times} 10^{-4}$ \cite{Couratin12}. In addition to the sudden change of the central potential, these calculations included corrections for the effect of the nuclear recoil of the daughter and for the direct collision mechanism between the beta particle and the electron. Both corrections were found to contribute by less than 1\%  to the ionization yield. For two-electron systems, the beta decay of neutral $^{6}$He atoms was also investigated experimentally as well as theoretically. The pioneering experimental results from Carlson et al. \cite{Carlson63} were first compared to the calculations of Wauters and Vaeck \cite{Wauters96}, and more recently to those of Schulhof and Drake \cite{Schulhof15}. Both calculations take into account electronic correlations and Auger emission and were found in good agreement with the experimental data for the single ionization probability. On the other hand, both calculations overestimate by about one order of magnitude the doubly ionization rate of 0.042(7)\% obtained by Carlson. This disagreement for the doubly ionization rate, pointing towards an  inaccuracy of the theory, has been recently confirmed by a new experiment using trapped $^{6}$He atoms \cite{Hong17}. These results already illustrate the difficulty for an accurate theoretical treatment of the electron SO process in a system comprising two electrons. 
 
 For multi-electron systems with more than two electrons, experimental results obtained in the $\beta^-$ decay of a collection of radioactive rare gas \cite{Carlson63Ar} have been compared to self-consistent-field calculations \cite{Carlson68} to test theoretical predictions for electron SO from the inner shell and from the outermost shell. This work, however, did not allow a precise test of the theory since it did not discuss the complete charge distribution and did not include the Auger emission subsequent to the primary SO ionization. Only recently, a few measurements were performed for $\beta^+$ decaying atoms of  $^{38m}$K \cite{Gorelov00}, $^{21}$Na \cite{Scielzo03} and $^{35}$Ar$^+$ singly charged ions \cite{Couratin13}. The experimental charge distributions obtained in the decay of  $^{21}$Na and $^{35}$Ar$^+$ were confronted to theoretical calculations based on the SA and including the recoil contribution as well as Auger emission \cite{Scielzo03,Couratin13}. For $^{21}$Na, a reasonable agreement between theory and experiment was obtained, although populations for charge-states greater than two were systematically overestimated by the calculations. In the case of $^{35}$Ar$^+$ decay, the calculations were found in excellent agreement with the experimental data. A strong contribution of Auger emission was evidenced, and the ionization reaction routes leading to the formation of all charge-states were identified. To obtain a more complete $Z$-dependent picture of the underlying ionization mechanisms, the study of other singly charged $\beta^+$ emitters is strongly required.

We present here our latest results obtained in the $\beta^+$ decay of $^{19}$Ne$^+$ ions. The $^{19}$Ne nucleus is a $\beta^+$ emitter whose decay to the $^{19}$F ground state is pure at 99.988\%, with a Q~value of 3238.4 keV. Its half life of 17.22~s is however somewhat challenging, yielding a low decay rate of 0.04~s$^{-1}$ per available parent nucleus.  
The experiment was performed at GANIL using the same technique and setup as for the $^{35}$Ar$^+$ decay study\cite{Couratin13}.  This setup, to which we refer to as LPCTrap \cite{Rodriguez06,Flechard08,Lienard15}, is based on the use of a Paul trap, to confine radioactive ions, coupled to a recoil-ion spectrometer giving access to both the energy and charge-state of the recoiling daughter. 
 %%%%%%%%%%%%%%%%%%%%%%%%%%%%%%%%%%%%%%%%%%%%%%%%%
\maketitle
\section{Experiment}
\label{Sec_Experiment}
\begin{figure}[!ht]
\includegraphics[width=85mm]{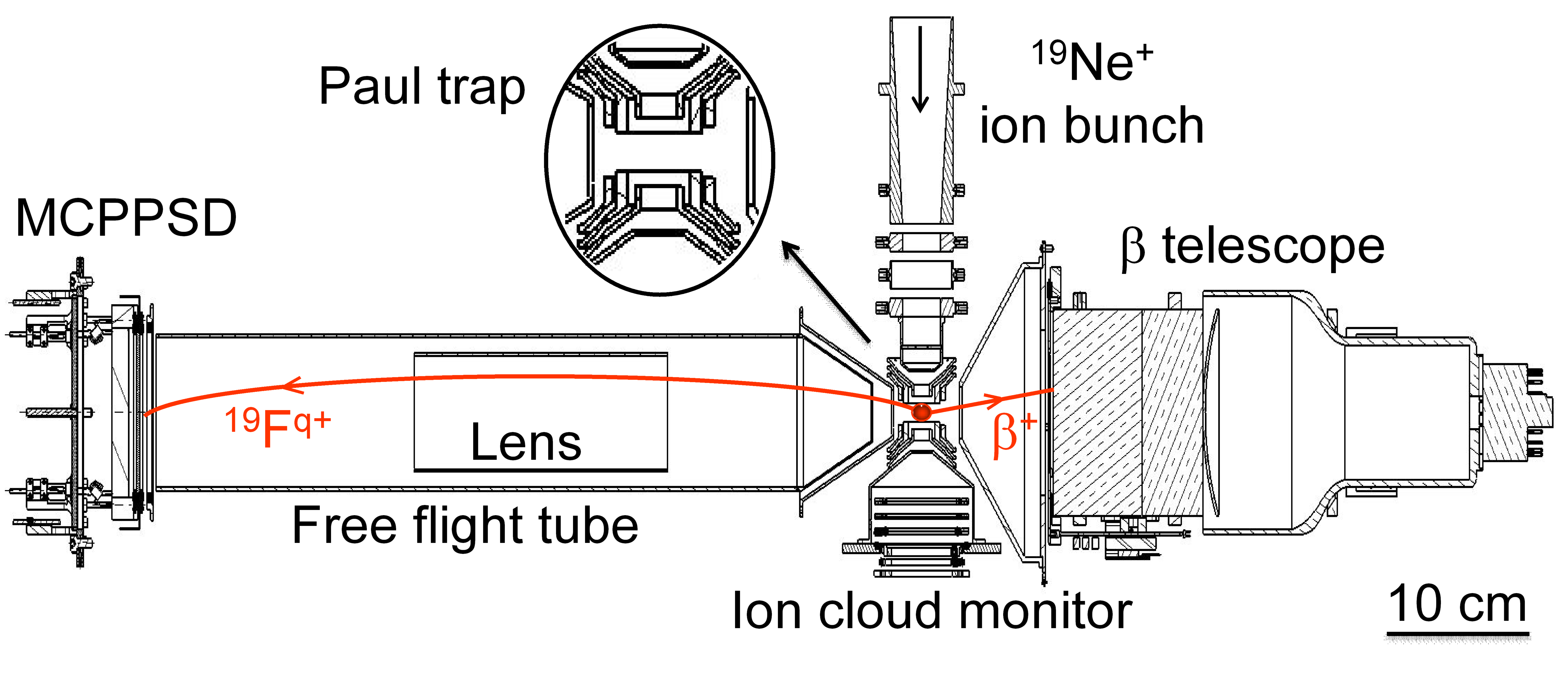}
\caption{Top view of the experimental setup. The insert shows the structure
of the six stainless steel rings of the Paul trap. See text for details.}
\label{fig.1}
\end{figure}
A general and technical description of the LPCTrap setup can be found elsewhere \cite{Flechard08,Flechard11,Couratin13,Ban13,Rodriguez06}. We detail only here the detection system and analysis method for experiments dedicated to shakeoff measurements.

 %%%%%%%%%%%%%%%%%%%%%%%%%%%%%%%%%%%%%%%%%%%%%%%%%
\maketitle
\subsection{Beam production and manipulation}
\label{Sec_beam}
The radioactive $^{19}$Ne nuclei were produced at the SPIRAL target-ECR ion source system \cite{Villari07} of GANIL, Caen, France, using a primary beam of  $^{20}$Ne ions at 95 MeV/nucleon impinging on a thick graphite target. The $^{19}$Ne production through projectile fragmentation was then optimal, with a relative contribution of unstable isobars estimated below $10^{-4}$ and whose effects are negligible at the present level of precision. The elements ionized by the ECR source were accelerated by the 9950~V potential of the source platform,  mass selected by a magnetic dipole of $m/q\sim$250 resolving power, and guided towards the LPCtrap setup through the LIRAT beamline.  At the exit of the ECR source, a strong contamination with a charge over mass ratio $q/m = 1/19$ prevented the direct use of singly charged $^{19}$Ne$^+$ ions. The contamination was, for the very dominant part, attributed to the presence of H$_3$O$^+$ and $^{18}$OH$^+$ molecular ions resulting in 30~nA of unwanted beam. Such a high current would indeed saturate the buncher used for beam preparation and injection in the measurement Paul trap. By selecting the beam with a charge over mass ratio $q/m = 2/19$, this contamination was suppressed due to molecular fragmentation within the ECR source plasma. Previous test experiments with stable  $^{20}$Ne had shown that this choice also resulted in a loss of about 60\% of the extracted Ne ions. During the run with radioactive ions, a beam with charge over mass ratio $q/m = 2/19$ of 150~pA to 200~pA containing approximately 40\% of  $^{19}$Ne$^{2+}$ ions and 60\% of $^{19}$F$^{2+}$ stable contaminants was continuously delivered to the LPCTrap experiment. The amount of $^{19}$Ne$^{2+}$ radioactive ions was monitored by counting $\beta$ particles on a removable Passivated Implanted Planar Silicon (PIPS) detector intercepting the beam at the entrance of the apparatus. A typical $^{19}$Ne$^{2+}$ beam intensity of $2\times10^8$~pps was deduced from this measurement. 

The ions were continuously injected in a Radio Frequency Cooler and Buncher (RFQCB) \cite{Darius04} for beam preparation. This is a 50 cm long linear Paul trap mounted on a high voltage platform to decelerate the ions down to 50 eV. The RFQCB is filled with He buffer gas at a pressure of $8\times10^{-3}$ mbar to cool down the ions below 1 eV.  The RF voltage of the RFQCB was chosen to confine ions with $q/m = 1/19$. About 30\% of the $^{19}$Ne$^{2+}$ ions injected in the RFQCB were converted in $^{19}$Ne$^{+}$ by charge exchange with the buffer gas, cooled by elastic collisions, and accumulated into bunches near the exit of the structure. Due to unstable trajectories of ions with charge over mass ratio $q/m = 2/19$, the remaining 70\% were lost by collisions on the electrodes and walls of the RFQCB. $^{19}$Ne$^{+}$ bunches were then extracted using a cycle period of 200~ms with a total transmission of the RFQCB of $\sim$4\%. They were re-accelerated downstream using a pulsed cavity, transported between the two traps with a kinetic energy of about 1~keV, and decelerated down to $\sim$100 eV by a second pulsed cavity located at the entrance of the measurement transparent Paul trap (MTPT). 

The MTPT is a 3-D Paul trap made of six concentric rings (Fig. \ref{fig.1}). $^{19}$Ne$^{+}$ ions were confined in the trap by applying a 1.55~MHz RF voltage of 120~V$_{pp}$ to the two inner rings. Two other rings were used to slow down or accelerate the ions during the injection and extraction phases by applying pulses of a few 100~V. Their potential was set to 0~V during the trapping period. The two external rings allowed the fine tuning of the trapping potential to optimize the trapping time of the ions in the trap. For each injection cycle, an average of about $1.7 \times 10^4$ $^{19}$Ne$^+$ ions were trapped and confined in the MTPT during 160~ms for data taking. Trapped ions were then extracted towards the ion cloud monitor (Fig. \ref{fig.1}) which is a micro-channel plate detector allowing the estimation of the amount of trapped ions. During the last 40~ms of each cycle, the trap was left empty for background data taking. Helium buffer gas at a pressure of $10^{-5}$ mbar was also used in the MTPT chamber to further cool down the trapped ions within the first 20~ms of the cycles.

 %%%%%%%%%%%%%%%%%%%%%%%%%%%%%%%%%%%%%%%%%%%%%%%%%
\maketitle
\subsection{Detection setup and data analysis}
\label{Sec_det}
The $\beta$ particles and the  $^{19}$F$^{q+}$ recoiling ions resulting from $\beta$ decay of trapped $^{19}$Ne$^+$ ions were detected by two detectors located around the trap (Fig. \ref{fig.1}). The $\beta$ telescope is composed of a thin double sided silicon strip detector followed by a plastic scintillator. The $60\times60$~mm strip detector provides the position of the incoming $\beta$ particles with a 1~mm resolution and allows the rejection of $\gamma$ rays triggering only the plastic scintillator. The latter gives the energy of the $\beta$ and also defines the reference time for a decay event. A recoil ion spectrometer enables the separation of the charge-states of recoiling ions from their TOF. Ions emitted towards the recoil ion spectrometer first cross a collimator through a 90\% transmission grid (set at ground potential) and are then accelerated by a $-2$~kV potential applied to a second 90\% transmission grid at the entrance of a 58~cm free flight tube (Fig. \ref{fig.1}). Inside the tube, an electrostatic lens at $-250$~V allows focusing the ions towards the center of a micro-channel plate position sensitive detector (MCPPSD) \cite{Lienard05} located at the end of the spectrometer. This detector comprises another 90\% transmission grid set at the same potential as the free flight tube and located 6~mm upstream from the set of two micro-channel plates. A $-4$~kV voltage applied on the front plate of the MCPPSD ensures a maximum  and uniform quantum efficiency for all charge-states of the recoil ions, independently of their initial kinetic energy. The absolute detection efficiency previously estimated in such conditions was found to be (52.3$\pm$0.3)\% due to the open area ratio of the MCP~\cite{Lienard05}.

\begin{figure}[!ht]
\includegraphics[width=85mm]{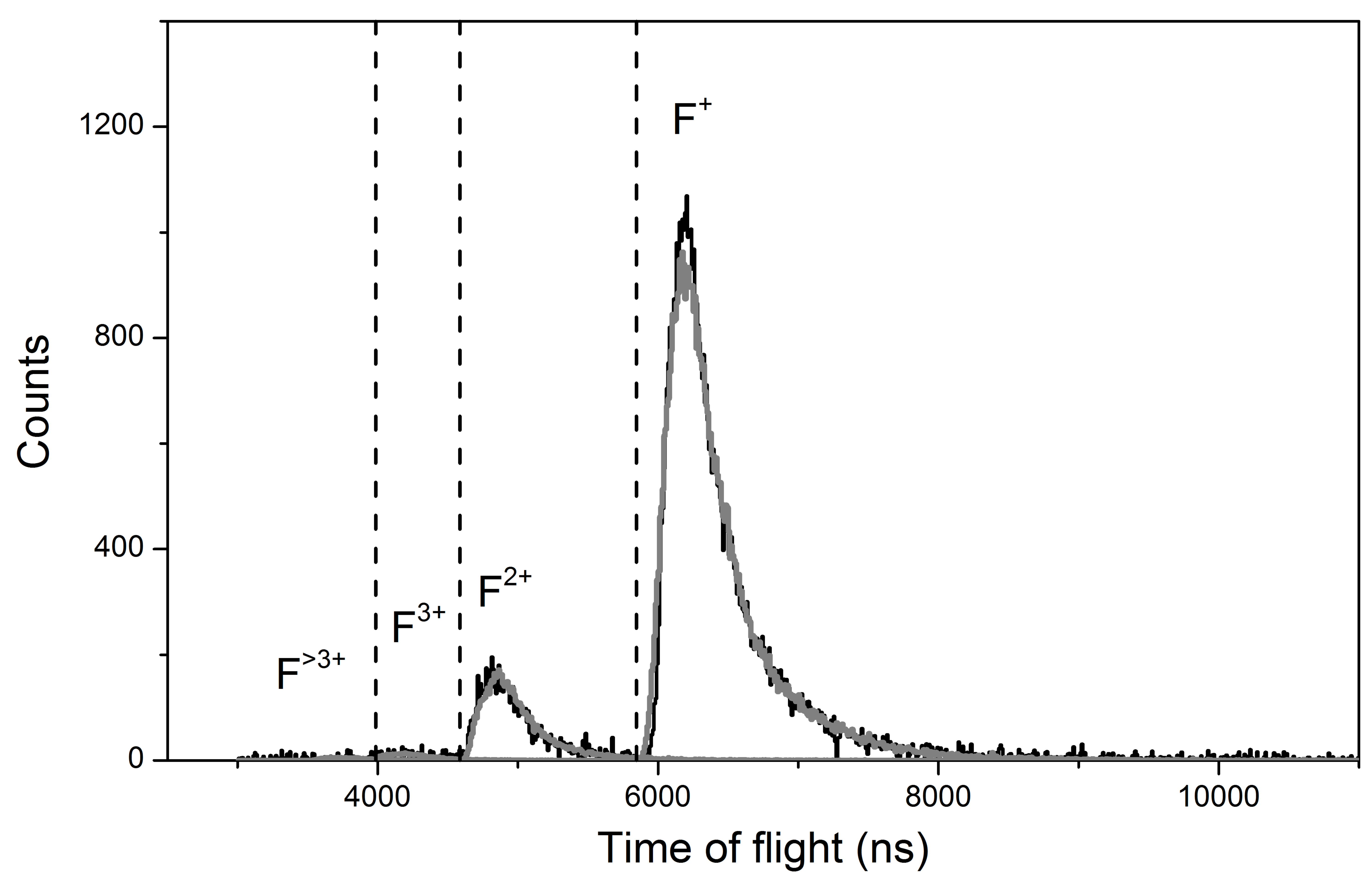}
\caption{Experimental (black line)  and simulated TOF spectra associated to the different charge-states (gray lines). The bin width of the histogram is 10~ns.
Vertical dashed lines indicate the ranges of integration used to obtain the charge-state branching ratios.}
\label{fig.2}
\end{figure}

For each detected event, the energy and position of the $\beta$ particle as well as the TOF and position of the recoil ion were recorded. The procedure applied for the detectors calibrations was identical to that described in Ref. \cite{Flechard11}. Only events corresponding to a $\beta$ particle depositing more than 0.4~MeV in the scintillator were kept in the analysis. A time reference within each 200~ms cycle was also sent to the acquisition in order to identify events recorded during the 40~ms period of background data taking and events recorded during the first 20~ms, prior the end of the cooling process. Data with time stamps between 20~ms and 160~ms were kept as ``good events''. Background events recorded between 160~ms and 200~ms correspond to decays from untrapped $^{19}$Ne. These events were used to correct for the contribution of untrapped $^{19}$Ne mixed with the good events in the 20-160~ms selection window. They represent less than 1\% of the good events and the systematic error associated to this correction is negligible. Another source of background arises from uncorrelated signals from the recoil ion and $\beta$ detectors. For a large part, these events are due to He buffer gas atoms triggering the MCPPSD, at an average rate of $\sim1500$~s$^{-1}$, in coincidence with $\beta$ particles hitting the $\beta$ telescope. This contribution yields a constant background in TOF corresponding to about 28~counts per 10~ns bin which was subtracted in the final TOF distribution shown in Fig.\ref{fig.2}, where the peaks associated to the different charge-states of $^{19}$F$^{q+}$ can be clearly identified.
\begin{figure}[!ht]
\includegraphics[width=85mm]{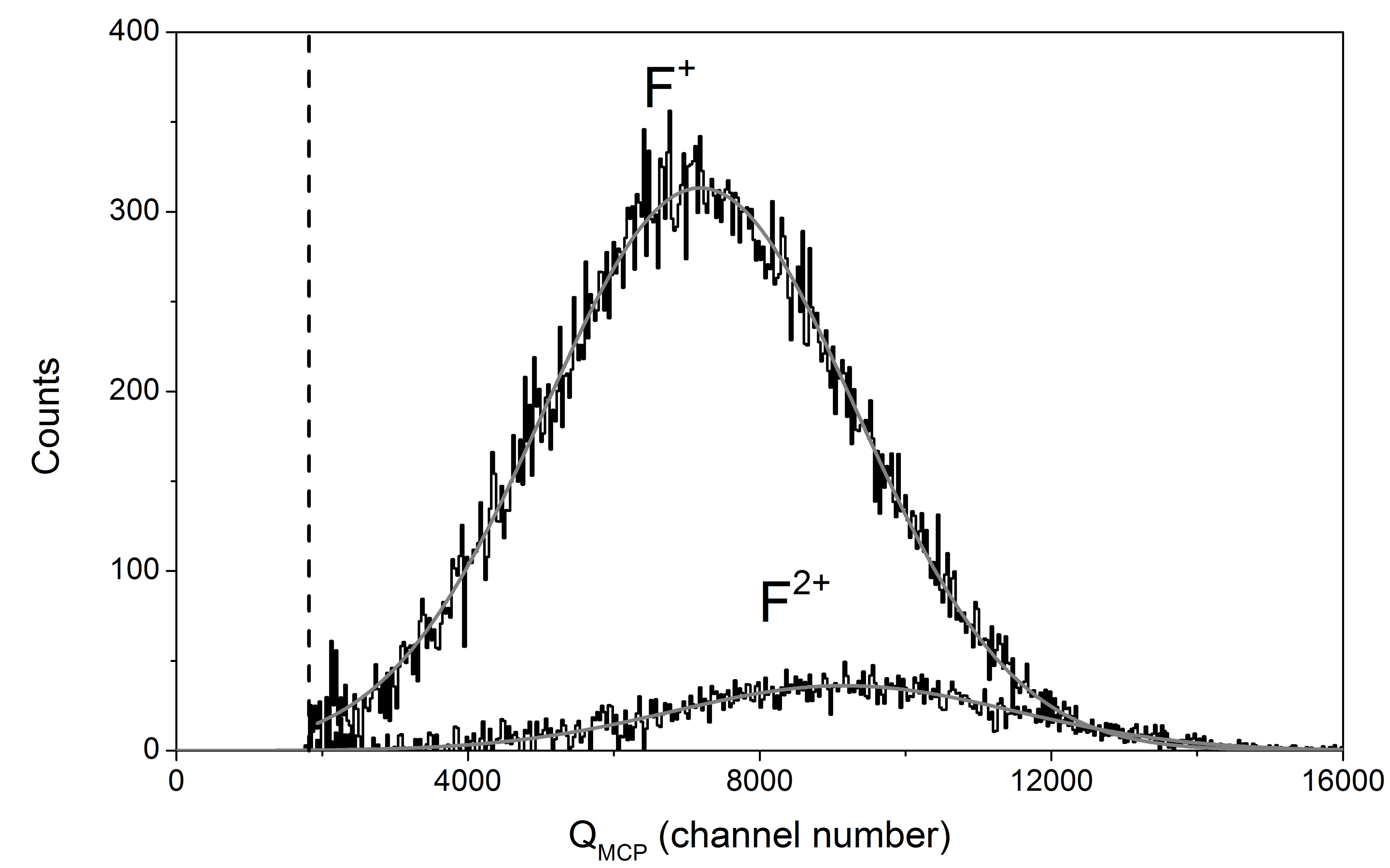}
\caption{Charge collected from the  MCPPSD for F$^{+}$ and  F$^{2+}$ recoil ion charge-states (black lines)  fitted with gaussians (gray lines). 
The vertical dashed line indicates the cut due to electronic threshold.}
\label{fig.3}
\end{figure}

The experimental charge-state branching ratios and their associated statistical uncertainty were simply deduced from the integration of counts within the TOF selection windows displayed on Fig.\ref{fig.2}. Background subtraction was accounted for in the evaluation of the statistical error. An additional correction, labeled Tail$_{\rm{corr.}}$ in Table I, takes into account the tails of charge distributions extending beyond their respective integration windows. In order to estimate properly this correction, the TOF spectra associated to each charge-state were generated using Monte-Carlo simulations~\cite{Flechard11,Couratin12} and were fitted to the experimental data. Several components of the simulations, such as the exact size of the trapped ion cloud and possible scattering of the $\beta$ particles on parts of the trapping chamber were neglected or approximated. Therefore, a conservative relative uncertainty of 10~\% was considered for these corrections.

Charge exchange with the He buffer gas is another process that could bias the charge-state ratio measurement. Electron capture probabilities from He buffer gas between the center of  the Paul trap and the entrance of the spectrometer have been estimated using experimental cross sections measured for Ne$^{q+}$ + He and  Ar$^{q+}$ + He collisions in the same velocity regime~\cite{Justiniano84}. Ionization potentials of fluorine being similar to those of neon and argon, the charge exchange cross sections for Ne$^{q+}$ and Ar$^{q+}$ ions constitute a good approximation of what one would expect with F$^{q+}$ ions. The highest charge-state involved here ($q$=4) yields the largest charge exchange probability which is below 10$^{-4}$. Electron capture processes can therefore be neglected at the present level of precision.

Another source of systematic effect could arise from an unequal detection efficiency for recoil-ions with different charge-states. The geometrical detection efficiency associated to the different charge-states was first investigated using the Monte-Carlo simulations. It was demonstrated that, well within statistical uncertainties, the same fraction of recoiling ions with charge-states $q$=+1 to $q$=+4 were collected on the surface of the MCP. This ruled out a possible effect of the trapping electric field on the detection solid angle associated to different charge-states. Then, the response function of the MCPPSD to the different populations of ions was studied. The initial kinetic energy of the recoil-ions ranges from a few~eV to 202~eV, independently of the charge-state. However, prior to hitting the surface of the micro-channel plate, the ions are accelerated by a $-4$~kV potential and higher charge-states gain more energy.  For ions with lower charge-states, this results in a lower pulse height of the MCPPSD signals and therefore a lower detection efficiency. The loss of detected events due to the electronic threshold was carefully estimated by fitting the charge distributions collected from the recoil ion detector with gaussian functions (Fig. \ref{fig.3}). It was found to represent only 0.55$\pm$0.05\% of events for F$^{+}$, and negligible for F$^{q+}$ with $q \geq 2$. This small correction labeled MCP$_{\rm{corr.}}$ was taken into account. Ideally, one could also account for a very small dependence of the MCP quantum efficiency on the ion charge state. Due to the large mean number of secondary electrons emitted by each ion when impinging the MCP, this second-order correction is expected to be smaller than the one due to the electronic threshold and was therefore neglected. For charged recoil ions, the experimental charge-state branching ratios including corrections are given in Table I. The systematic error associated to the corrections Tail$_{\rm{corr.}}$ and MCP$_{\rm{corr.}}$ were found negligible compared to the statistical error given with the experimental results.
\begin{table}[!htb]%[H] add [H] placement to break table across pages
\caption{Experimental ion charge-state branching ratios after corrections and details of the corrections (\%) }
\label{tab:results}
\begin{ruledtabular}
\begin{tabular}{cccc}
Charge &  Exp. & MCP$_{\rm{corr.}}$ & Tail$_{\rm{corr.}}$  \\
 & results &  \\
\hline
1 & 87.6 $\pm$0.6 & 0.1 & -0.2  \\
2 & 11.8  $\pm$0.3 & -0.1 & +0.2  \\
3 & 0.6 $\pm$0.2 & 0.0 & 0.0  \\
$\ge$4 & 0.0 $\pm$0.2 & 0.0 & 0.0  \\
\end{tabular}
\end{ruledtabular}
\end{table}

For a dominant part of the decay events, there is no electron shakeoff. The $\beta$$^+$ decay of a $^{19}$Ne$^+$ ions then results in recoiling neutral $^{19}$F atoms, that are insensitive to the electric field of the recoil-ion spectrometer and to the post-acceleration field of the MCPPSD. The detection probability is thus much smaller than for recoil-ions because of the smaller collection solid angle and of the very low intrinsic detection efficiency of the MCPPSD for atoms with energy ranging from 0 to 202~eV. This resulted in a very limited statistics for these events, with TOF always larger than 13 $\mu$s. They do not appear on the spectrum shown in the Fig. \ref{fig.2} and are well separated from the ions contribution. 

The MCPPSD response function being very difficult to qualify for such low energy atoms, we chose to use the number of $\beta$ particles detected in singles (without condition on the detection of a recoil) to estimate the fraction of the decays leading to the production of a neutral recoiling atom. Knowing the overall absolute detection efficiency for ions, the fraction of  singles events associated to charged $^{19}$F$^{q+}$ recoils can be inferred, the rest being associated to neutral $^{19}$F atoms. For decay events from trapped radioactive $^{19}$Ne$^+$, the experimental ratio between the number of detected $\beta$~-~recoil-ion coincidences and the number of detected singles events was found to be R$_{\rm{exp}}=(2.88\pm0.09)\times10^{-2}$. The uncertainty is dominated here by the subtraction procedure of the untrapped decay contribution. Using the Monte Carlo simulation and by considering that all the decays yield charged recoil-ions, the corresponding ratio was found to be R$_{\rm{sim}}=(9.5\pm1.0)\times10^{-2}$, where the uncertainty is now dominated by the absolute efficiency of the MCPPSD for ions, conservatively estimated as $(52\pm3)$\%, and by the transparency of the three $(90\pm2)$\% transmission grids located on their way. The difference between these two ratios arises from the fraction of coincidence events with neutral $^{19}$F atoms that are not detected. We can thus infer the fraction of charged ions, R$_{\rm{exp}}/$R$_{\rm{sim}}=30.5\pm4.2\%$, and of neutral atoms, $69.5\pm4.2\%$, resulting from the $\beta$ decay process.
 %%%%%%%%%%%%%%%%%%%%%%%%%%%%%%%%%%%%%%%%%%%%%%%%%

\maketitle
\section{Theoretical Calculations}
\label{Sec_calculations}
The theoretical framework of our calculations concerning the $^{19}$F$^{q+}$ and $^{35}$Cl$^{q+}$ charge-state distributions which result from 
the $\beta$ decay of $^{19}$Ne$^+$ and $^{35}$Ar$^+$, respectively, is presented here below. 

\subsection{Shakeoff ionization}
The nuclear decay of the parent ion $^A$X$^+$ leads to the appearance of daugther ionic species $^A$Y$^{q+}$ through shakeoff ionization as a result of the 
non-adiabatic rearrangement of the electron cloud as the $^A$X nucleus transforms into $^A$Y. In the present work, $^A$X stands for either $^{19}$Ne or 
$^{35}$Ar, and $^A$Y$\equiv ^{19}$F or $^{35}$Cl, respectively. The multi-electron dynamics of $^A$X$^+$ are described using the Independent Particle Model (IPM) 
which assumes that each electron $i$ of the system evolves independently from the others, subject to the mean field created by the nucleus and the remaining electrons 
(see \cite{Bransden} and references therein). Therefore the IPM probability to ionize $q_S$ electrons among the $N$ total ones of $^A$X$^+$ is 
\begin{equation}
P_{q_S}^{ion}=\sum_{i_1=1}^{N}p_{i_1} \sum_{i_2>i_1}^{N}p_{i_2} \dots \sum_{i_{q_S}>i_{q_S-1}}^{N}p_{i_{q_S}} \prod_{j\neq i_1,\dots,i_{q_S}}^{N}{\left ( 1 - p_{j} \right )}
\label{pSO}
\end{equation}
where $p_{i}$ is the one-electron ionization probability for the $i^{th}$ electron. The IPM probability is expected to be accurate when the electron-electron interaction
is small compared with the interaction between the electrons and the nucleus, which generally occurs in ionization from inner shells of systems having a 
large nuclear charge $Z$. 

Neglecting shakeup processes in the sudden rearrangement of the electron cloud, which would correspond to electron transitions into bound states of $^A$Y, the one-electron 
probability $p_i$ is defined by
\begin{equation}
p_i=1- \sum_{n'\le n'_{max}}{| < \varphi^{(\text{Y})}_{n'l} \text{e}^{\text{i} {\bf K.r}} |  \varphi^{(\text{X}^+)}_{n_il_i} > |^2}
\label{pi}
\end{equation}
in the rest frame of $^A$Y of mass $M$, which recoils with the energy $E_R$ and associated wavevector $K=\sqrt{2E_R/M}$ (in atomic units). 
$\varphi^{(\text{X}^+,\text{Y})}_{nl}$ are the electron wavefunctions describing one electron orbiting in the $nl$ subshell of X$^+$ and Y, respectively,
and $n'_{max}$ is the principal quantum number of the outermost shell of X$^+$ ($n'_{max}=2$ for X=Ne and $n'_{max}=3$ for X=Ar). Since $K$ is small, the mean recoil energy can be used instead of integrating over the recoil energy distribution and $\text{e}^{\text{i} {\bf K.r}}$ can be expanded in eq.(\ref{pi}) to obtain the alternative expression
\begin{eqnarray}
p_i&=&1- \sum_{n'\le n'_{max}} |< \varphi^{(\text{Y})}_{n'l_i} | \varphi^{(\text{X}^+)}_{n_il_i} > |^2 \nonumber \\
&+&K^2 | < \varphi^{(\text{Y})}_{n'l_i \pm 1} | {\bf r} |  \varphi^{(\text{X}^+)}_{n_il_i} > |^2 \\
&-&K^2 \text{Re} (<\varphi^{(\text{Y})}_{n'l_i}|\varphi^{(\text{X}^+)}_{n_il_i}>^\ast < \varphi^{(\text{Y})}_{n'l_i} | \text{r}^2 | \varphi^{(\text{X}^+)}_{n_il_i} >) \nonumber
\label{dev}
\end{eqnarray}
up to second order in $K$. This expression shows that ionization stems from the coherent superposition of static orbital mismatch, in terms of the
$< \varphi^{(\text{Y})}_{n'l_i} | \varphi^{(\text{X}^+)}_{n_il_i} >$ overlaps, and recoil effects, through the $K^2$-dependent terms. 

\begin{table*}
\caption{One-electron ionization probabities $p_i$ as functions of the subshell to which pertains electron $i$ in F$^+$ recoiling with 
energy $E_R=161$ eV. The probabilities are obtained by means of UHF or ROHF approaches, with triple-$\zeta$ aug-cc-pVTZ and enlarged 
quadruple-$\zeta$ aug-cc-pVQZ underlying gaussian basis \cite{Dunning89}.}
\begin{tabular}{c|ccc}
\hline
\hline
 subshell   & $1s$ &  $2s$   & $2p$ \\
\hline
UHF/aug-cc-pVTZ & 6.981$\times 10^{-3}$ & 2.921$\times 10^{-2}$ & 3.794$\times 10^{-2}$ \\
ROHF/aug-cc-pVTZ & 7.015$\times 10^{-3}$ & 2.910$\times 10^{-2}$ & 3.757$\times 10^{-2}$ \\
UHF/aug-cc-pVQZ & 6.981$\times 10^{-3}$ & 2.929$\times 10^{-2}$ & 3.816$\times 10^{-2}$ \\
ROHF/aug-cc-pVQZ & 7.014$\times 10^{-3}$ & 2.918$\times 10^{-2}$ & 3.777$\times 10^{-2}$ \\
\hline
\hline
\end{tabular}
\label{HF}
\end{table*}

In practice, the $\varphi^{(\text{X}^+,\text{Y})}_{nl}$ orbitals entering eq.(3)
%(\ref{dev})
are obtained by means of Hartree-Fock (HF) calculations using the GAMESS-US 
quantum chemistry package \cite{Gamess}. Since we deal with open-shell systems, we can either implement restricted open-shell (ROHF) or unrestricted (UHF) methodologies. 
As shown in Table \ref{HF} for the case of Ne$^+$ where $E_R=161$ eV (the mean energy of the detected recoil ions) so that $K=0.018 42$ atomic units, both methods lead to almost identical results for the one-electron
probabilities $p_i$. We also prove in Table \ref{HF} the convergence of our results with respect to an increase of the size of the underlying gaussian basis employed in the quantum chemistry calculations, since large quadruple-$\zeta$ calculations yield results in close agreement with those issued from smaller triple-$\zeta$ computations.
%These large-scale calculations finally confirm the reliability of previous HF computations of $\varphi^{(\text{Ne}^+,\text{F})}_{nl}$ by 
%Clementi and Roetti \cite{Clementi74} which, when introduced in eq.(\ref{dev}), yield $p_i$ probabilities very close to ours. 
The results presented in the next section are based on the one-electron probabilities $p_i$ obtained by means of the UHF approach with the largest underlying gaussian basis. 

\subsection{Subsequent Auger processes}
SO leads to ionic species $^A$Y$^{q_S^+}$ which can present vacancies in their inner shells. These excited ions then relax, by means of either decaying radiative transitions 
or Auger processes. While the former do not change the ionic charge-state, the Auger decay involves the ejection of electrons and thus yields significantly higher
charge-states. Note that a single vacancy can lead to the ejection of several electrons through a so-called Auger cascade such as, for instance, Cl$^+$($1s2s^22p^63s^23p^5$) 
$\rightarrow$ Cl$^{2+}$($1s^22s2p^63s^23p^4$)+$e$ $\rightarrow$ Cl$^{3+}$($1s^22s^22p^63s^23p^2$)+$2e$. 
High-order Auger decay is also involved in cases where primary SO ionization
leads to multiple inner-shell vacancies. We take into account all these additional ionization mechanisms within our IPM treatment using the Auger probabilities 
$\tilde{p}^{s,m_i}_i$ of Ref. \cite{Kaastra93}. $\tilde{p}^{s,m_i}_i$ corresponds to the probability for Auger emission of $m_i$ electrons after the inner shell 
electron $i$ has been removed from $^A$Y$^{s+}$. On the basis of the multi-electron SO probability (\ref{pSO}), the final probability for SO emission of $q_S$ electrons 
followed by the ejection of $q_A$ Auger electrons is then defined as 
\begin{eqnarray}
P_{q_S,q_A} & = & \sum_{m_{i_1},\dots,m_{i_{q_S}} \atop m_{i_1}+\dots+m_{i_{q_S}}=q_A}\sum_{i_1=1}^{N}{p_{i_1} \tilde{p}_{i_1}^{s,m_{i_1}}} \sum_{i_2>i_1}^{N}{p_{i_2} \tilde{p}_{i_2}^{s,m_{i_2}}} \dots  \nonumber \\
&  & \sum_{i_{q_S}>i_{q_S-1}}^{N}p_{i_{q_S}} \tilde{p}_{i_{q_S}}^{s,m_{i_{q_S}}} \prod_{j\neq i_1,\dots,i_{q_S}}^{N}{\left ( 1 - p_j \right )}.
\label{pfinal}
\end{eqnarray}
The probability to form the daughter charge-state $q$ corresponds to the sum of all $P_{q_S,q_A}$ such that $q=q_S+q_A$. Therefore, the contributions 
of primary SO and subsequent Auger processes can be disentangled for fixed $q$. Further, the nature and multiplicity of both primary SO vacancies and related Auger decays 
are encoded in the IPM formulation (\ref{pfinal}) of each $P_{q_S,q_A}$. In other words, all the ionization routes leading to the charge-state $q$ can be identified and their
relative weights can be easily determined. 

 %%%%%%%%%%%%%%%%%%%%%%%%%%%%%%%%%%%%%%%%%%%%%%%%%
\maketitle
\section{Discussion}
\label{Sec_discussion}
Experimental and theoretical charge-state distributions for $^{35}$Cl$^{q+}$ have been compared in \cite{Couratin13}. We then mainly focus here on  
the $^{19}$F$^{q+}$ results. However, comparing the features of the $^{19}$F$^{q+}$ and $^{35}$Cl$^{q+}$ distributions enables to obtain a $Z$-dependent picture of the 
underlying ionization mechanisms and to gauge the capabilities and limitations of our theoretical treatment. 

{\setlength{\tabcolsep}{0.8em}
\begin{table*}
\caption{Experimental $^{19}$F$^{q+}$ ion charge-state relative branching ratios (\%) compared to calculations with and without recoil and Auger ionizations.}
%\begin{tabularx}{\textwidth}{cccccc}
\begin{tabular}{cccccc}
\hline
\hline
            & Exp.    &  With recoil & Without recoil & With recoil   & Without recoil \\
 Charge $q$ & results &  With Auger   & With Auger     & Without Auger & Without Auger \\
\hline
1 &        87.6 $\pm$0.6  & 84.25 & 84.29 & 88.75 & 88.83 \\
2 &        11.8 $\pm$0.3  & 13.84 & 13.80 & 10.51 & 10.44 \\
3 &        0.6  $\pm$0.2  & 1.74  & 1.73  & 0.71  & 0.70 \\
$\geq 4$ & 0.0  $\pm$0.2  & 0.17  & 0.17  & 0.03  & 0.03 \\
\hline
\hline
\end{tabular}
\label{Table-ratios}
\end{table*}

\subsection{$^{19}$F$^{q+}$ ion charge-state distribution}
We compare in Table \ref{Table-ratios} the experimental charge-state branching ratios with their theoretical counterparts. The results from full calculations, including Auger decay 
and recoil effects, are found in reasonable agreement with the measurements for the main $q=1,2$ charge-states but they do not fall within the experimental error bars. Larger
discrepancy is observed for the higher charge-states that the calculations significantly overestimate (by about a factor three in the case of $q=3$). As mentioned in Sec.~\ref{Sec_Intro}, 
theoretical overestimation of high charge-state populations has already been observed in the $\beta$ decay of $^6$He \cite{Carlson63,Wauters96,Schulhof15,Hong17} and $^{21}$Na \cite{Scielzo03}. 
While electron-electron correlations were explicitly included in the two-electron $^6$He case, the theoretical framework employed for $^{21}$Na was similar to the present IPM one. 
On the other hand, our approach has yielded very good agreement with experiment for all the charge-states in the case of $^{35}$Ar$^{+}$ decay \cite{Couratin13}. These observations 
would lead one to conclude that the agreement for $^{35}$Ar$^{+}$ is fortuitious. 
This is not the case: we have explained in the previous section that the IPM, coupled to the underlying mean-field (HF) description of one-electron transitions, 
is expected to be accurate for large-$Z$ systems where the electron-electron correlations, which entangle the electron dynamics, are small compared to the
electron-nucleus interaction. This is why the IPM and related HF treatment, based on independent electron transitions, makes a much better job for Ar$^+$ than for Ne$^+$ and
all other low-$Z$ systems. 

The (relative) inaccuracy of the IPM also shows up in the total ionization probability associated to the decay of $^{19}$Ne$^{+}$. This latter has been estimated 
to 30.5 $\pm$ 4.2$\%$ experimentally and the IPM yields 23.5$\%$. It is important to note that this probability does not depend on Auger processes which only
redistribute the ionization flux over the $q > 1$ charge-states once one or several inner shell vacancies have been created by SO ionization. The liability 
of the experimental/theoretical discrepancy is therefore related to the mean-field computation of the ionization probabilities $P^{ion}_{q_S}$. In this respect, 
taking into account the 
shakeup processes that have been neglected in the present work would increase even more the discrepancy since it would lead to a decrease of the ionization probabilities 
$p_i$, which would yield an even smaller total ionization probability. Once again, a much better agreement is obtained in the case of the higher-$Z$ $^{35}$Ar$^+$ 
system whose decay leads to 72 $\pm$ 10$\%$ of $^{35}$Cl atoms in the experiment and to 73.9$\%$ in the calculations. 

One can take advantage of the simple IPM formulation of Eq. (\ref{pfinal}) to estimate the relative importance
of the ionization mechanisms that contribute to the formation of the charge-states $^{19}$F$^{q+}$. Recoil ionization effects can be artificially canceled by setting 
$K=0$ in the one-electron probabilities of Eq. (3). It is clear from Table \ref{Table-ratios} that the charge-state distribution is almost unaffected by the recoil: primary SO 
ionization mostly consists of pure static orbital mismatch. Auger processes can be neglected in order to observe the charge-state distribution which results only from SO; 
the $^{19}$F$^{q+}$ populations are then computed by means of Eq. (\ref{pSO}) with $q_S=q$ (setting $K=0$ in Eq. (3) or not). As it may be expected, the $q=1$ charge-state
branching ratio then increases while the higher ones decrease (see Table \ref{Table-ratios}). It seems therefore that we reach a better agreement with the experimental values. 
However this is accidental since Auger decay has to occur for sure when a vacancy is created by SO in the $1s$ shell of fluorine. The probability to fill such a $1s$ vacancy, 
with simultaneous emission of one electron, is taken from Ref. \cite{Kaastra93} and has a value of $\sim$1. This is expected, and considered accurate, for such a rather small 
$Z$-system where the Auger decay from the L ($n=2$) shell to the K ($n=1$) one is known to completely dominate the radiative deexcitation \cite{Venugopala72}. 
Therefore, it is plausible that the $1s$ $p_i$ probability for primary ionization is overestimated, leading to an excessive Auger redistribution of the ionization flux 
into high charge-states. In this respect, decreasing artificially the $1s$ probability by 30$\%$ allows to obtain a good agreement of the full calculations with experiment. 
However such a reduction is too large to be interpreted as the contribution of non-ionizing shakeup transitions from the inner shell, and 
this does not fix the problem of the underestimated total ionization probability.

\subsection{Importance of Auger decay versus $Z$}
We illustrate in the upper panel of Fig. \ref{fig.4} the comparison of the experimental F$^{q+}$ branching ratios with the results of our calculations. Beyond the misleading
impression that Auger emission should be suppressed to make the agreement with experiment better, the Auger processes do not drastically change the whole ionic distribution. 
The $q=1$ ratio decreases by $\sim$4.5$\%$ and this amount is shared between the higher charge-states. We understand such a moderate change since the only way for Auger 
processes to occur in fluorine is to empty the inner $1s$ shell whose ionization probability is small (see Table \ref{HF}). In contrast, very significant changes
appear in the Cl$^{q+}$ distribution when Auger ionization is allowed (see the lower panel of Fig. \ref{fig.4}). In this case, primary SO is not able to yield
$q \geq$ 5 states whose population is entirely driven by Auger emission, and the magnitude of the lower $q=4$ channel increases by more than one order
of magnitude when Auger emission is introduced in the theory. This increasing importance of the Auger decay is simply due to the existence of 
the supplementary M ($n=3$) shell in chlorine, which leads not only to the onset of additional Auger transitions as vacancies are produced in the L shell by primary SO 
but also to a higher multiplicity of the Auger cascades related to the K-shell vacancies. Observing the drop-off of Auger transitions would require considering 
higher-$Z$ species: the average number of electrons emitted during the decay of inner shell vacancies decreases for $Z > 25$ \cite{Kaastra93}, and  
the fluorescence then starts to override the non-radiative process. This means that the highest charge-state should be observed around $Z=25$, and the daughter 
charge-state distribution should shrink towards low $q$-values for heavier species. 

\begin{table*}
\caption{Main ionization routes leading to F$^{q+}$ formation (in \%). $nl^{-1}$ refers to primary SO hole creation in the $nl$-subshell of F
while $m \times e_A$ means emission of $m$ Auger electrons}
\begin{tabular}{cccc}
\hline
\hline
F$^+$ & F$^{2+}$ & F$^{3+}$ & F$^{4+}$ \\
\hline
$2p^{-1}$ : \textbf{76.62}  & $2s^{-1}2p^{-1}$ : \textbf{38.15} & $1s^{-1}2p^{-1}+1e_A$ : \textbf{51.40} & $1s^{-1}2p^{-2}+1e_A$ : \textbf{42.79}  \\
$2s^{-1}$ : \textbf{23.31} & $2p^{-2}$ : \textbf{37.02} & $2s^{-1}2p^{-2}$ : \textbf{17.77} & $1s^{-1}2s^{-1}2p^{-1}+1e_A$ : \textbf{32.55} \\
 & $1s^{-1}+1e_A$ : \textbf{32.57} & $1s^{-1}2s^{-1}+1e_A$ : \textbf{15.64} & $1s^{-2}+2e_A$ : \textbf{9.41}  \\
 & $2s^{-2}$ : \textbf{2.14} & $2p^{-3}$ : \textbf{11.68} & $2s^{-1}2p^{-3}$ : \textbf{7.40}  \\
 & & $2s^{-2}2p^{-1}$ : \textbf{3.38} & $2s^{-2}2p^{-2}$ : \textbf{2.81}  \\
 & & & $1s^{-1}2s^{-2}+1e_A$ : \textbf{2.48}\\
 & & & $2p^{-4}$ : \textbf{2.43} \\
\hline
\hline
\end{tabular}
\label{Table-routes}
\end{table*}
\begin{figure}[!ht]
\includegraphics[width=85mm]{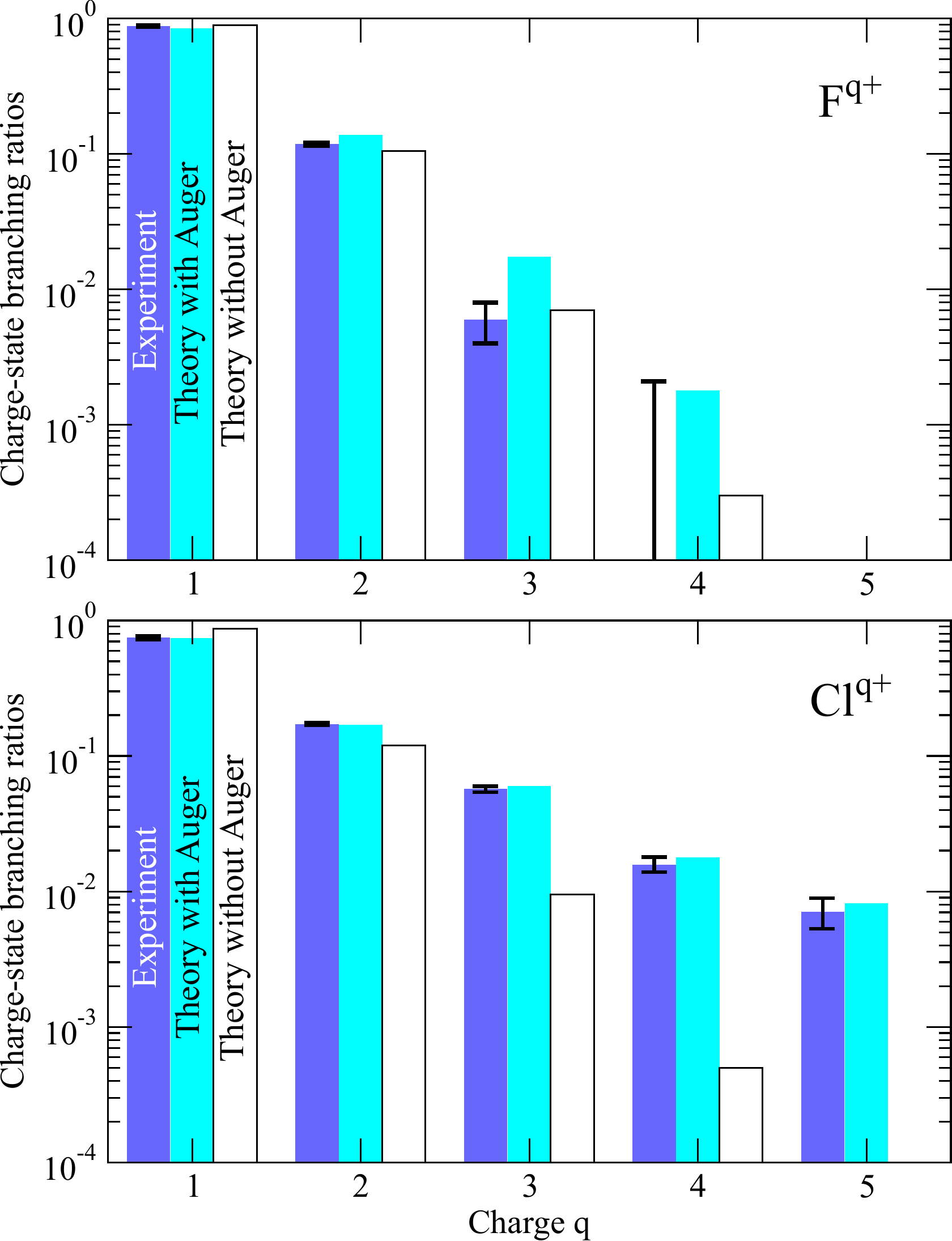}
\caption{Experimental and calculated charge-state branching ratios for $^{19}$F$^{q+}$ and $^{35}$Cl$^{q+}$ production subsequent 
to the $\beta$ decay of $^{19}$Ne$^{+}$ and $^{35}$Ar$^{+}$, respectively.}
\label{fig.4}
\end{figure}

\subsection{Ionization routes from $^{19}$Ne$^+$ to $^{19}$F$^{q+}$}
Table \ref{Table-routes} lists the ionization pathways which contribute more than 1$\%$ to the formation of the $^{19}$F$^{q+}$ charge-states
after the nuclear decay of $^{19}$Ne$^+$. As mentioned above, this identification is made by unrolling the combinatory formulation (\ref{pfinal}) of
the probabilities $P_{q_S,q_A}$ to form the charge-state $q=q_S+q_A$. As expected, single SO ionization involves one of the valence electrons, with a 
leading contribution ($\sim$77$\%$) of the $2p$ electron because of its smaller ionization potential. 
Auger emission starts contributing to the double ionization channel, with a relative weight of $\sim$33$\%$, but two-electron SO ionization 
from the L-shell is dominant for $q=2$. Single Auger processes, following double SO ionization, constitute two thirds of the $q=3$ channel. Finally, double Auger emission 
appears for $q=4$, as it may be expected since it requires to vacate totally the K-shell beforehand.

 %%%%%%%%%%%%%%%%%%%%%%%%%%%%%%%%%%%%%%%%%%%%%%%%%
\maketitle
\section{Conclusion}
\label{Sec_conclusion}
We have presented measurements of the charge-state distribution $^{19}$F$^{q+}$ resulting from the $\beta$ decay of $^{19}$Ne$^{+}$. 
Small experimental uncertainties have been obtained in spite of the challenging long half life of the $^{19}$Ne nucleus, owing to the optimized LPCtrap 
setup. 

Calculations based on the Independent Particle Model, which uses one-electron probabilities from the mean-field Hartree-Fock approach, have been 
performed to compare the charge-state distribution with the experiment. The comparison does not display the excellent consistency found previously for $^{35}$Cl$^{q+}$. We have traced back the root of the theoretical shortcomings to the IPM which does not provide accurate enough ionization probabilities for systems with low nuclear charge. In such systems, the electron-electron correlations can not be neglected with 
respect to the electron-nucleus interactions so that the independance of the electron dynamics, assumed in the IPM, is not valid. Furthermore, 
the Auger decay is not as important in fluorine as it is in chlorine so that the primary shakeoff probabilities are there of crucial importance. 
In chlorine, the production of high charge-states is controlled by Auger cascades related to vacancies in the K- and L-shells that the IPM describes 
better because of the higher $Z$. 

The ionization processes underlying the charge-state distribution resulting from $\beta$ decay should therefore be described by theoretical treatments 
which explicitely include the electron-electron interaction. This is not an easy task, even for He for which recent calculations did not succeed in yielding 
an accurate double ionization rate \cite{Schulhof15}. Such calculations for fluorine will constitute a theoretical challenge, as the difficulty 
increases further for higher-Z systems because of the larger multiplicity of coupled continua related to the available inner-shell vacancies.

\begin{acknowledgments}
The authors thank the LPC staff for their strong involvement in the
LPCTrap project and the GANIL staff for the preparation of a high quality ion beam.
\end{acknowledgments}
%%%%%
%\bibliography{biblio}
%
%%%%%%%%%%%%%%%%%%%%%%%%%%%%%%%%%%%%%%%%%%%%%%%%%%%%%

%
\end{document}